\documentclass[aps,reprint,amsmath,amssymb,superscriptaddress,floatfix,twocolumn
,nofootinbib
,prl
]{revtex4-2}
\usepackage{natbib,pmboxdraw}
\usepackage{graphicx,dcolumn,lipsum,placeins}
\usepackage[utf8]{inputenc}
\usepackage[T1]{fontenc}
\usepackage[all]{nowidow}
\usepackage{physics}
\usepackage[bottom]{footmisc}
\usepackage[mathlines]{lineno}
\usepackage[dvipsnames,table,xcdraw]{xcolor}
\usepackage[space]{grffile}
\usepackage[linesnumbered,ruled,vlined]{algorithm2e}
\DeclareUnicodeCharacter{25A0}{
	\rule[1pt]{5.25pt}{3pt}}
\usepackage{hyperref}[backref]
\hypersetup{
	colorlinks = true,
	allcolors = {blue},
}

\usepackage{appendix}
\usepackage{orcidlink}
\begin{document}
\def\r{.95}
\def\rt{.95}
\def\txt{Successful rate of 40,000 experiments on qubit flipping using different \mbox{C-Not} gates setups on a specific IBM QC.}
\title{Hacking quantum computers with row hammer attack}
\author{Fernando Almaguer-Angeles \orcidlink{0000-0002-3521-7256}}
\email{fernando.almaguerangeles@phdstud.ug.edu.pl}
 \affiliation{International Centre for Theory of Quantum Technologies, University of Gdańsk, Jana Bażyńskiego 1A, 80-309 Gdańsk, Poland}
 \author{Pedro R. Dieguez \orcidlink{0000-0002-8286-2645}}
\email{pedro.dieguez@ug.edu.pl}
\affiliation{International Centre for Theory of Quantum Technologies, University of Gdańsk, Jana Bażyńskiego 1A, 80-309 Gdańsk, Poland}
\author{Akshata Shenoy H. \orcidlink{0000-0002-6703-8383}}
\email{akshata.shenoy@ug.edu.pl}
 \affiliation{International Centre for Theory of Quantum Technologies, University of Gdańsk, Jana Bażyńskiego 1A, 80-309 Gdańsk, Poland}
\author{Marcin Pawłowski \orcidlink{0000-0002-8611-947X}}%
 \email{marcin.pawlowski@ug.edu.pl}
\affiliation{International Centre for Theory of Quantum Technologies, University of Gdańsk, Jana Bażyńskiego 1A, 80-309 Gdańsk, Poland}
\date{\today}

\begin{abstract} 
We demonstrate a hardware vulnerability in quantum computing systems by exploiting \mbox{cross-talk} effects on an available commercial quantum computer~(IBM). Specifically, based on the \mbox{cross-talk} produced by certain quantum gates, we implement a row hammer attack that ultimately allows us to flip a qubit. Both \mbox{single-qubit} and \mbox{two-qubit} operations are performed and analyzed. Our findings reveal that \mbox{two-qubit} operations applied near the target qubit significantly influence it through \mbox{cross-talk}, effectively compromising its state.
\end{abstract}

\maketitle

Quantum computing is advancing towards \mbox{real-world} applications~\cite{Preskill2018quantumcomputingin,abughanem2024nisq}, promising unprecedented processing power~\cite{arute2019quantum} with potential breakthroughs in fields such as cryptography~\cite{portmann2022security}, material science~\cite{lordi2021advances,bauer2020quantum}, and artificial intelligence~\cite{park2024trends}. However, as these systems become more integrated into infrastructures involving sensitive data processing, security concerns grow~\cite{fitzsimons2017private,PhysRevLett.111.230501,PhysRevLett.111.230502}.
Modern computing systems rely on confidentiality, integrity, availability, and safety of critical processes~\cite{samonas2014cia,ward2019security,Biasin2022}, making it essential to assess and mitigate potential vulnerabilities.  In fact, quantum computers~(QC) are vulnerable to integrity attacks~\cite{pino2020demonstration,abraham2013cross,piltz2014trapped}. QCs are susceptible to \textit{\mbox{cross-talk}}, a phenomenon in which unintended interactions between the nearest neighboring qubits can compromise their computational accuracy and reliability. Understanding and mitigating these vulnerabilities is essential in determining whether quantum computing can be trusted for critical applications.

In quantum computing, \mbox{cross-talk} occurs when one component of an experimental setup—such as a qubit, control line, electromagnetic field, resonator, or photodetector—unintentionally influences another~\cite{sarovar2020detecting}. These unintended interactions can introduce errors, compromise computational accuracy, and, more critically, be exploited as a security vulnerability.
Here, we explore the quantum counterpart of the \mbox{well-known} \textit{row hammer attack}, a hardware vulnerability commonly exploited in classical computing. In classical systems, a row hammer attack involves repeatedly accessing specific memory rows to induce bit flips in adjacent memory locations. These unwanted \mbox{bit-flip} errors in neighboring rows, known as target rows, increase with the number of devices per area on a hardware unit. Row hammer attacks have been attempted with repetitive access to single, two, or multiple targets~\cite{kim2014flipping}. It is known that refreshing the target rows repetitively provides a simple yet inefficient solution to the issue. There exists other complex \mbox{software-based} countermeasures as well though all of them do not entirely solve the problem~\cite{kim2024rowhammer}.

\begin{figure}
  \centering
  \includegraphics[width=.9\linewidth]{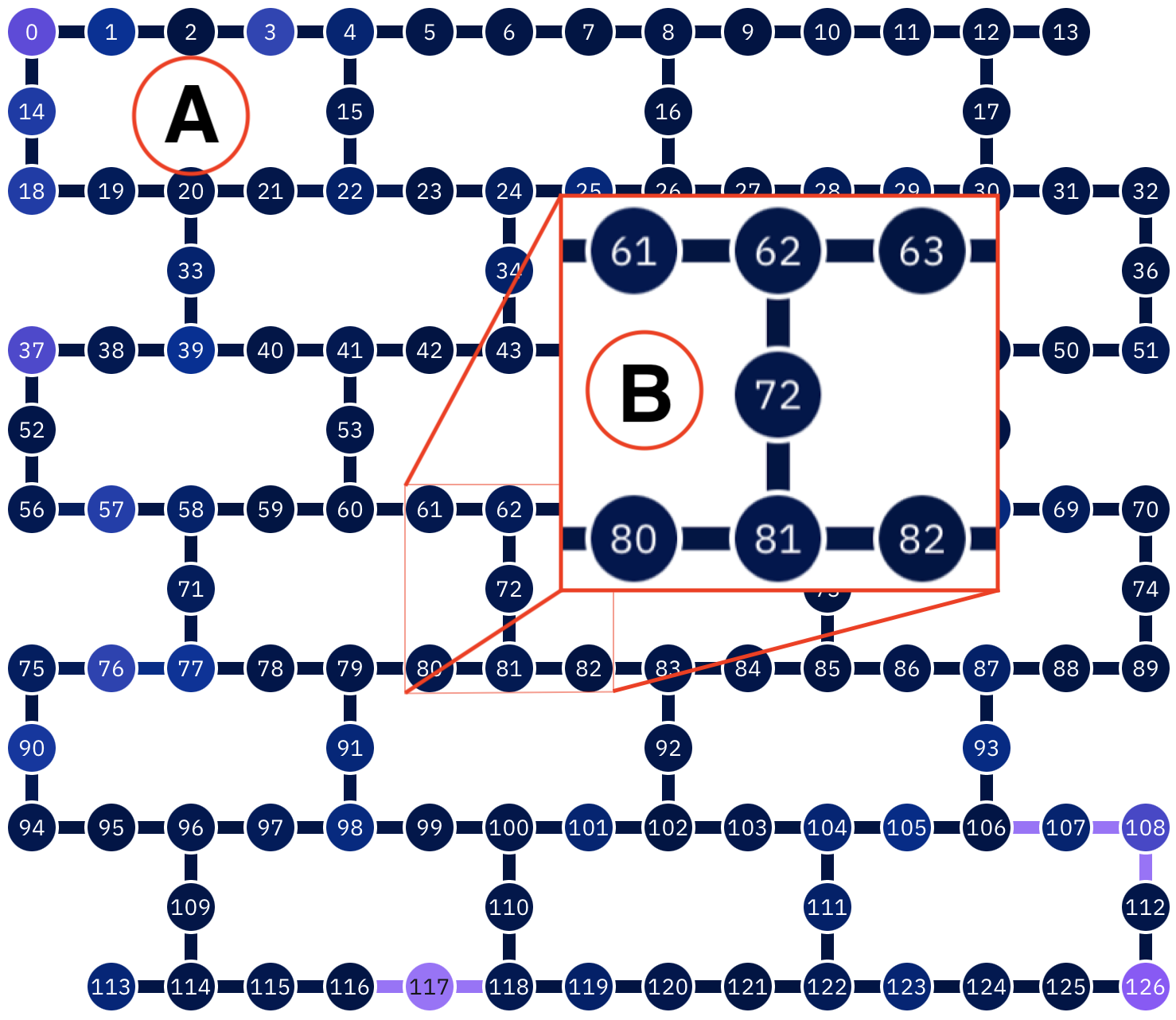}
  \caption{A: IBM's QC qubit topology.
  B: Example of the structure to be used in the experiments.
  Grid of qubits, represented by numbered circles, and their physical connection, represented by a line connecting two qubits in a topological arrangement of 127 qubits.}
  \label{fig:topo}
\end{figure}

Row hammer attacks have been used to escalate privileges and execute system calls beyond a sandbox, allowing access to a limited subset of \mbox{x86-64} machine instructions~\cite{cve}. A report from the CVE Program~\cite{cve_2016}, a cybersecurity vulnerability database, described an incident in October 2016 where an Android application exploited row hammer, among other techniques, to gain root access on several popular smartphones.
The vulnerability stems from the physical limitations of the newer DRAM chips. In response, Google’s security research team disclosed a new row hammer exploit~\cite{Google_Online_Security_Blog_2021}, highlighting the ongoing risks associated with this attack. Security researchers have further demonstrated that row hammer exploits can be independent of both system architecture and instruction set~\cite{arXiv_1507_06955}.

In this letter, we demonstrate how to exploit \mbox{cross-talk} in IBM's QC architecture shown in Fig.~\ref{fig:topo} by designing a \mbox{row-hammer} attack to compromise it. Specifically, we investigate whether it is possible to flip a target qubit by applying a set of universal quantum gates~(\textbf{QG}).
To explore this vulnerability, we implement both \mbox{single-qubit} rotations and \mbox{two-qubit} operations. Focusing on \mbox{controlled-not}~(\textbf{\mbox{C-Not}}) gates, our results reveal that significant \mbox{cross-talk} can be induced allowing flipping of a qubit with high probability. To validate our approach, we conduct experiments on three different IBM QC, analyze their chip architecture, and design row hammer attacks. We provide a detailed algorithm for the attack ensuring reproducibility of quantum circuits. A Cramér’s V statistical analysis is also performed to assess its feasibility in establishing a cause and effect of using these gates to induce \mbox{crosstalk}. The results are divided into two main parts. The first verifies whether or not single qubit gates induce sufficient \mbox{cross-talk}. The second part explores \mbox{cross-talk} produced by a \mbox{two-qubit} gate such as the \mbox{C-Not}. Our findings demonstrate the potential security risks posed by \mbox{cross-talk} in QCs and highlight the need for improved mitigation strategies.

\textit{\underline{\smash{Experiments}}-}
The experiments were run in \textit{Brisbane}, \textit{Kyiv}, and \textit{
Sherbrooke} IBM's QCs, with a \mbox{10-minutes} \mbox{free-usage} monthly account, and Qiskit \textit{sampler} method.
All IBM's QCs used in this work have the same qubit topology, and can be seen in Fig.~\ref{fig:topo}~A. Certain Points were chosen to increase \mbox{cross-talk} and perform the experiments. These points followed Fig.~\ref{fig:topo}~B, where QG were put around the \textit{centre}. Giving us a total of 18 centres. For simplicity, we choose uniformly only six centres to collect data from to demonstrate the vulnerability. These are $15$, $34$, $54$, $72$, $93$, and $109$.

\begin{figure}[h!]
  \centering
  \includegraphics[width=\r\linewidth]{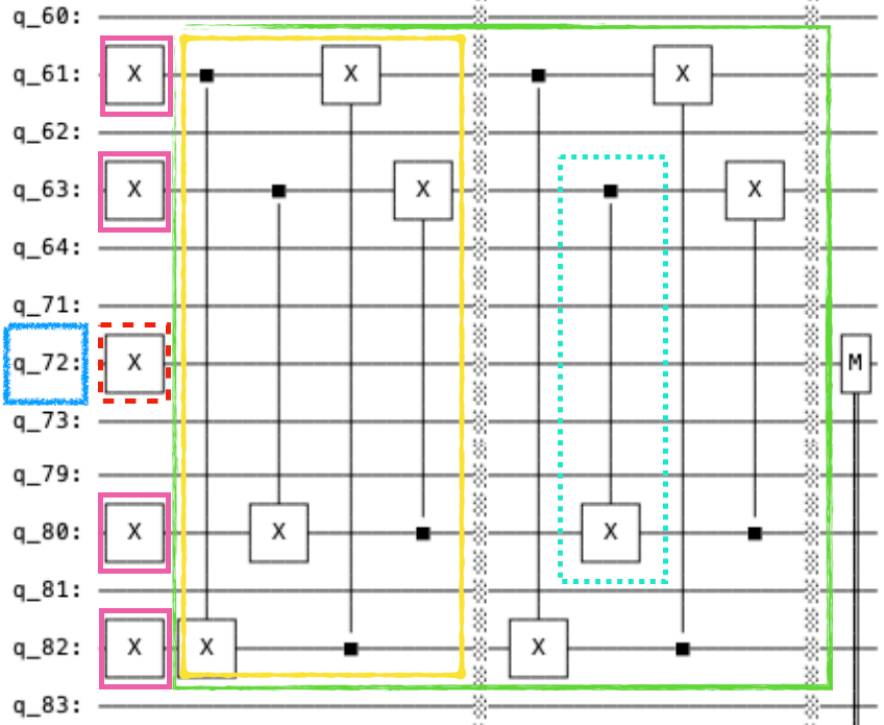}
  \caption{The Experiment: In the above circuit, the pink box indicates state preparation. The cyan box marks the gate operations that are involved. While the red dashed box shows the initial state of the measuring qubit, the blue box represents the target qubit to be flipped and the green box encompasses the number of additional gate sets applied. Finally, the Yellow box defines the base experiment setup, specifying the gate type and arrangement.}
  \label{fig:labelling}
\end{figure}

The experimental circuit is labeled as follows. In Fig.~\ref{fig:labelling}, the pink boxes indicate that the qubits have been initialized in the state $|1\rangle$ as opposed to the default $|0\rangle$ in IBM QC. The blue box indicates the target qubit $q_{72}$ which has been initialized in $|1\rangle$ as indicated by the red dashed box. The green box represents the \mbox{full-structure} gate that will be employed in the present case. The yellow box represents the primary gates that were used. The configuration depicted in Fig.~\ref{fig:labelling}, the target qubit is labeled \textcolor{pink}{-}\textcolor{ cyan}{cx}(\textcolor{red}{-}\textcolor{blue}{72}-\textcolor{green}{1}) \textcolor{yellow}{4} \textit{cross}\footnote{[pink]-[cyan]cx([red]-[blue]72-[green]1) [yellow]4 \textit{cross}} because the involved operation qubits form a cross in the chosen subtopology of qubits 61, 63, 80, and 82.

\begin{figure}[!h!p!]
  \centering
  \includegraphics[width=.65\linewidth]{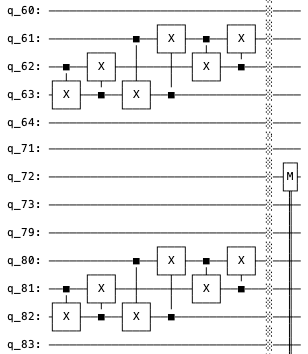}
  \caption{An example of a quantum circuit using a suitable subtopology in IBM's QC. The \mbox{set-up} demonstrates how a \mbox{C-Not} gate's \mbox{cross-talk} can be employed to orchestrate a row hammer attack and flip a qubit.}
  \label{fig:exp}
\end{figure}

The experiments are aimed to measure \mbox{cross-talk} on IBM QCs by performing various gate operations around a central qubit and measuring its output. An example of this setup is illustrated in Fig.~\ref{fig:exp}, where the central qubit is chosen to be $72$ which is surrounded by other qubits~(according to the topology $61$, $62$, $63$, $80$, $81$, and $82$), which were involved in the gate operations. Two types of gates were used in these experiments: \mbox{single-qubit} gates and the \mbox{two-qubit} \mbox{C-Not} gate.

For the \mbox{single-qubit} gate experiments, P, RZ, RX, RY, $\sqrt{x}$, $\sqrt{x}^\dagger$, Y, and U gates were chosen. The definition of these gates can be found in the Appendix~\hyperref[app:singlequbit]{A}. 
The experiments were conducted using different configurations, each with an increasing number of gates. Every experiment was repeated $40,000$ times, measuring only the output of the target~(central) qubit. Due to the complexity of the configurations and the large number of executions, specialized \mbox{factory-style} software was used to generate quantum circuits consistently in a reproducible manner. The algorithm behind this software is presented in Appendix~\hyperref[appx:algo]{B}.

Errors in IBM's QCs can be classified into two main categories. The first category involves hardware errors which include metrics such as error per layered gate, readout assignment error, probability of measuring $0$ when preparing $\ket{1}$, \mbox{Z-axis} rotation~(rz) error, $\sqrt{x}$~(sx) error, \mbox{Pauli-X} error, and ECR error, among others~\cite{sampler_2024}. The second category pertains to \mbox{software-induced} errors, particularly those associated with \mbox{Python-based} Qiskit.
IBM continuously monitors and recalibrates hardware metrics through its calibration data processes. To account for these fluctuations, a control experiment was included in each experimental round to assess the QC’s precision at that specific moment. This control data provides a baseline for output comparison, helping to identify the presence of \mbox{cross-talk}.
In contrast, the errors related to IBM's sampler method are \mbox{non-physical}, meaning they are not used to evaluate readout or gate performance. Instead, these errors manage the program's logic flow and provide feedback to software developers. For instance, they include errors such as \textit{invalid arguments given}, which are not tied to the hardware's physical behavior but ensure the proper execution of the quantum program ~\cite{sampler_2024}.

\textit{\underline{Results}-}
The results of the experiments are presented as follows. First, the data from the \mbox{single-qubit} experiments are used for the rest of the analysis.  Second, the most representative \mbox{two-qubit} setup experiment is explained, while experimental results using the other two  QCs are presented in Appendix~\hyperref[appx:results]{D}.

The collected data is stored in a CSV file similar to Table~\ref{tab:exp}. 
\begin{table}[!h]
  \centering
    \resizebox{\rt\linewidth}{!}{
\begin{tabular}{rcc}\rowcolor[HTML]{C0C0C0} 
\textbf{Experiment} & \textbf{Output 0 \%}        & \textbf{Output 1  \%} \\
Precision(93) 	 	   		 	& 97.2675					  & 2.7325    \\\rowcolor[HTML]{EFEFEF}
p(93-1000) 6    	   		 	& 97.365                      & 2.635     \\
rz(93-1000) 6      	 	     	& 97.365                      & 2.635     \\\rowcolor[HTML]{EFEFEF}
rx(93-1000) 6      		     	& 96.7325                     & 3.2675    \\
ry(93-1000) 6      			 	& 96.6425                     & 3.3575    \\\rowcolor[HTML]{EFEFEF}
$\sqrt{x}$(93-1000) 6   		 	& 96.9575                     & 3.0425    \\
$\sqrt{x}^\dagger$(93-1000) 6   & 96.995                      & 3.005     \\\rowcolor[HTML]{EFEFEF}
y(93-1000) 6        			& 96.245                      & 3.755     \\
u(93-1000) 6        			& 96.6875                     & 3.3125
\end{tabular}}
  \caption{Brisbane single-qubit experiments results.}
  \label{tab:exp}
\end{table}
This table illustrates the gate sets around the central qubit~(in this example, $1,000$ additional sets are included), denoted as qubit $93$ in the \textit{Experiment} column. The \textit{Output} columns represent the observables measured in these experiments. Since the experiments focused on a qubit not undergoing any operations, an Output 0 was expected. Increasing inaccuracy in expectation values between distant qubits reveals the level of noise present, as discussed in previous studies~\cite{ibmHelloWorld,ibmExactNoisy}.
 Therefore, as higher the percentage of Output 1, the higher of qubit flipping and success of the row hammer attack.

Table~\ref{tab:exp} demonstrates that the experiment is stable and does not induce \mbox{cross-talk} to adjacent channels. This stability is also observed in the results from the Kyiv quantum processors, details in Appendix~\hyperref[appx:fliprate]{C}. However, Sherbrooke is an exception, giving us more than 82.89\% of qubit flipping. Yet, these results are achieved at the cost of lower precision, which in this experiment is $83.35\%$. This indicates that the \mbox{single-qubit} method would not be effective for conducting a successful row hammer attack due to the number of \mbox{single-qubit} gates required.

\begin{figure}[!h!]
  \centering
  \includegraphics[width=\r\linewidth]{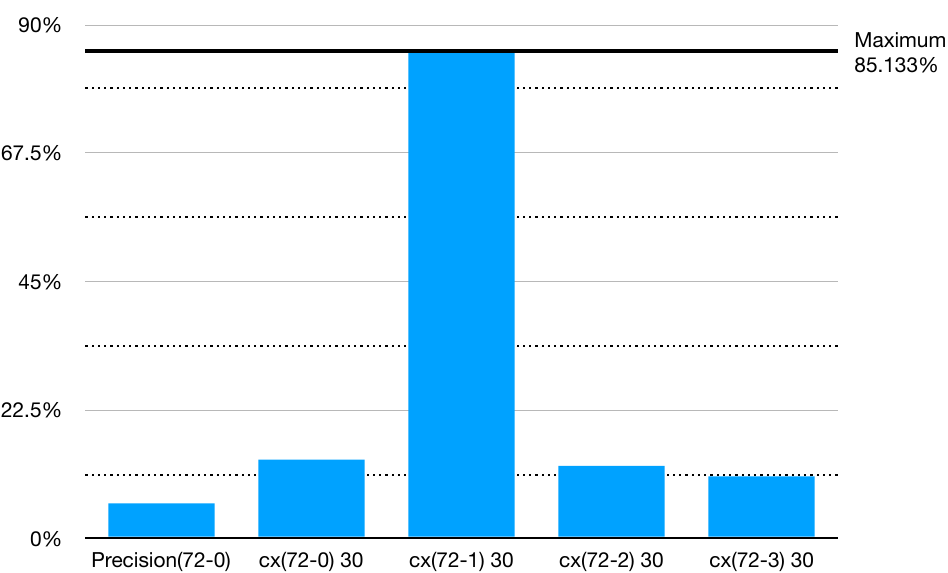}
  \caption{Sherbrooke A: Flipping $|0\rangle$ to $|1\rangle$.
 \txt}
  \label{fig:sherbrooke}
\end{figure}

\begin{figure}[!h!]
  \centering
  \includegraphics[width=\r\linewidth]{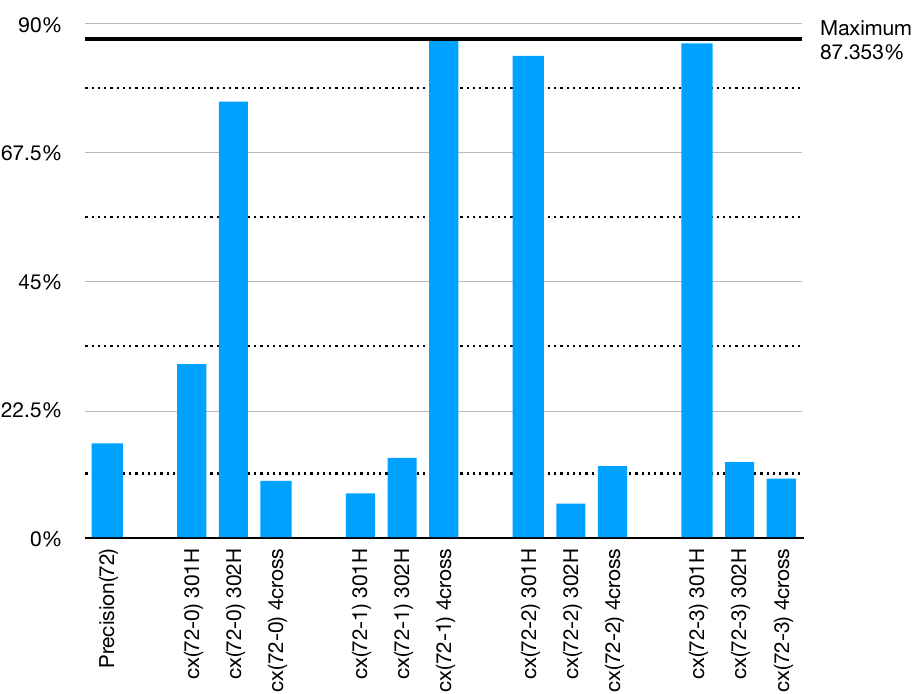}
  \caption{Sherbrooke B: Flipping the state $|0\rangle$ to the $|1\rangle$. \txt}
  \label{fig:sherbrookeBis}
\end{figure}

Sherbrooke's analysis, for the \mbox{C-Not} gate, includes two different objectives. On the one hand, in Figs.~\ref{fig:sherbrooke}~-~\ref{fig:sherbrookeBis}, the target qubits are being tried to be flipped from $|0\rangle$ to $|1\rangle$. Fig.~\ref{fig:sherbrooke}, depicts one such successful attack. It was performed on centre $72$ with $30$ \mbox{C-Not} configuration and one extra set, obtaining a bit flip for more than $85.13\%$. In Fig.~\ref{fig:sherbrookeBis}, we have four successful attacks. Configuration using $30$ \mbox{C-Not} gates around qubit $72$, two Hadamard gates on qubits $61$, and $63$, on the top qubit row, and in the same fashion on the bottom qubit row, and no extra sets, has a little less than $ 76.39\%$ of successful qubit flip. Another successful configuration is a \mbox{four-gate} cross configuration, an extra set of gates, around qubit $72$ that has slightly more than $87.35\%$ of qubit flip. The last successful configuration has two configurations, which is a $30$ \mbox{C-Not} gate configuration with two and three extra gate sets, one Hadamard gate at qubit $62$ on the top, and one at the bottom on qubit $81$. The first of these configurations has less than $84.35\%$ of qubit flip, while the second has more than $86.65\%$ of qubit flip. 

\begin{figure}[h!]
    \centering
    \includegraphics[width=\r\linewidth]{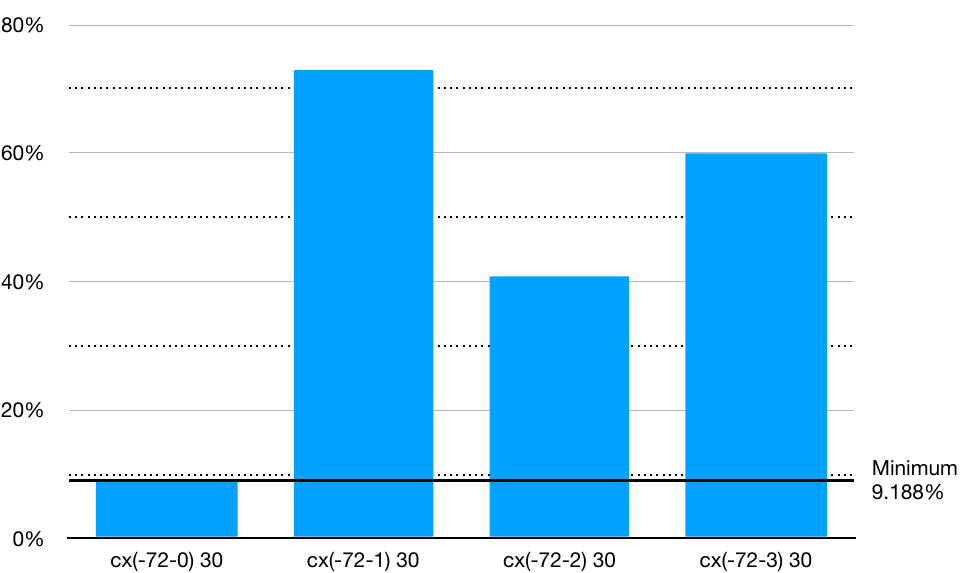}
    \caption{Sherbrooke: Flipping the state $|1\rangle$ to the $|0\rangle$. \txt}
    \label{fig:sherbrooke1}
\end{figure}

The results indicate that both configurations, with and without entanglement, led to successful attacks, making it unclear whether the entanglement produced by \mbox{C-Not} gates play a significant role in the attack. On the other hand, in Fig.~\ref{fig:sherbrooke1}, attempts to the flip qubit from $|1\rangle$ to $|0\rangle$, shows only one successful event for the cx$(-72-0)30$ configuration. It has a slightly more than $90.81\%$ of observing qubit flip. It is worth mentioning that these experiments have a precision of $93.83\%$ for Fig.~\ref{fig:sherbrooke}, while for the rest of the plots, above $83.31\% $. Therefore, these flipping qubit percentages occurred with a relatively high precision. Moreover, to determine whether these categorical variables are correlated, a Cramér's V analysis is detailed in Appendix~\hyperref[appx:cramers]{E}.

\begin{figure}[h!]
  \centering
  \includegraphics[width=\r\linewidth]{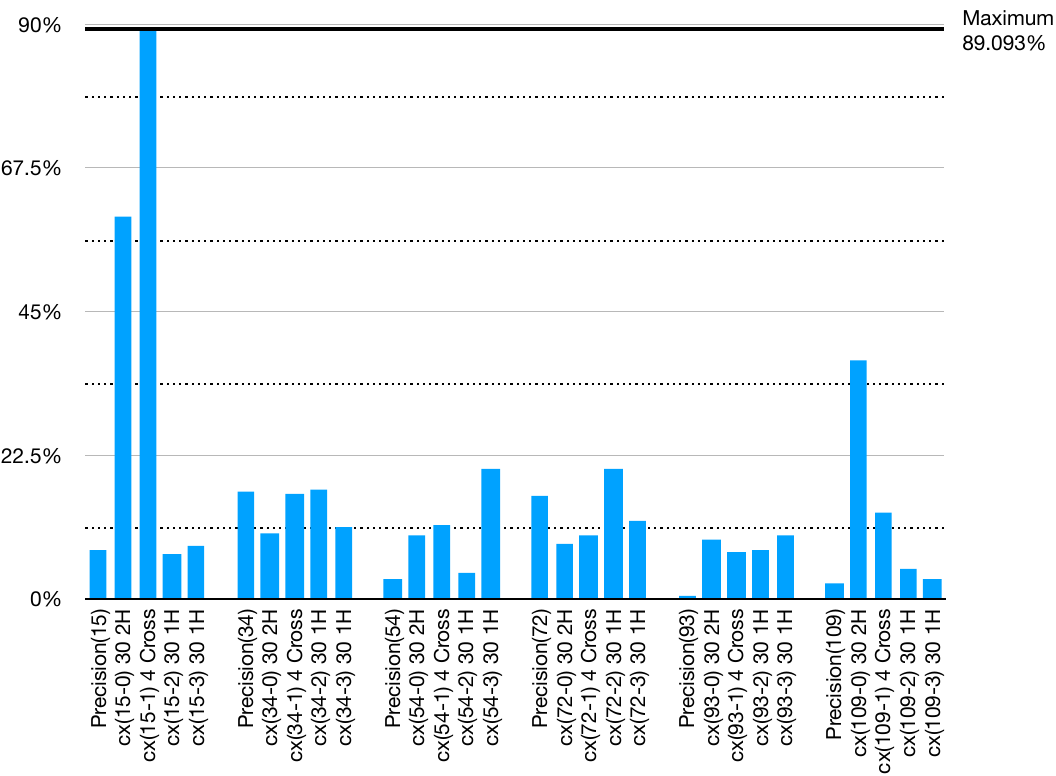}
  \caption{Sherbrooke: Flipping the state $|0\rangle$ to the $|1\rangle$ on different centres. \txt}
  \label{fig:sherbrookeCentres}
\end{figure}

\begin{figure}[h!]
  \centering
  \includegraphics[width=\r\linewidth]{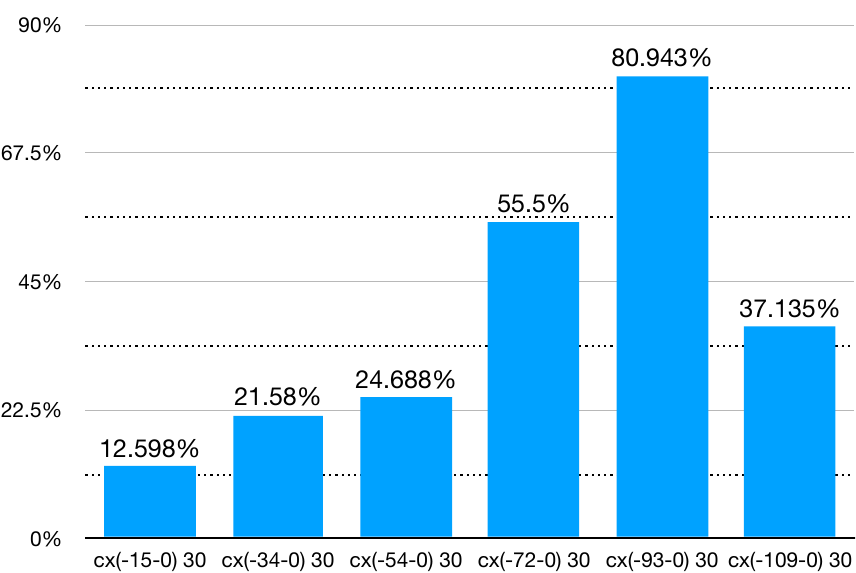}
  \caption{Sherbrooke: Flipping $|1\rangle$ to $|0\rangle$ on different centres. \txt}
  \label{fig:sherbrookeCentres1}
\end{figure}

The previous results led to further experiments, which were conducted at the previously selected centers using the successful attack settings. Figs.~\ref{fig:sherbrookeCentres}~-~\ref{fig:sherbrookeCentres1} show these centres and the flipping percentage on the different settings. Fig.~\ref{fig:sherbrookeCentres} shows only one successful attack. Configuration cx$(15-1)4$ cross has a qubit flipping rate of a bit more than $89.09\%$, and a precision of $92.44\%$. Fig.~\ref{fig:sherbrookeCentres1} shows three successful attacks. The first one, with slightly more than $87.40\%$ configuration cx(-15-0)30, has a precision of $92.44\%$. The second one is cx$(-34-0)30$ with $78.42\%$ of qubit flipping and a precision of more than $83.21\%$. The last one is cx$(-54-0)30$ with a successful qubit flipping rate of more than $75.31\%$, and a precision of a little over $96.73\%$.
The results for Kyiv and Brisbane QC are shown on Appendix~\hyperref[appx:results]{D}.

These experiments have been classified as row hammer attacks due to the repeated application of quantum operations to induce unintended state changes in the target qubit. Specifically, we executed $40,000$ operations on the target qubit, analogous to the repeated memory accesses in classical row hammer attacks. This persistent ``hammering'' exploited \mbox{cross-talk} effects to manipulate the target state. In the most effective cases, this method achieved a success rate of just under $94.83\%$, demonstrating the reliability of the attack in inducing controlled bit flips.

\textit{\underline{Discussion}-}
In this letter, we showed how to perform a successful row hammer attack on a QC by using certain QG and \mbox{cross-talk} to flip a qubit. This represents a significant security risk, as QCs are expected to handle sensitive applications, including cryptographic tasks and blind computation. The ability to flip a qubit undermines the reliability of quantum operations and raises concerns about the trustworthiness of quantum computations.  

To address this issue, further research is needed to fully understand the mechanisms behind \mbox{cross-talk-induced} attacks, develop reliable detection methods, and implement effective countermeasures. Future studies should explore mitigation strategies, such as optimized qubit layouts, \mbox{error-correcting} protocols, and \mbox{hardware-level} solutions to minimize \mbox{cross-talk} effects. As quantum technology advances, ensuring its security and resilience against such vulnerabilities will be crucial for its safe and reliable adoption in critical applications.

\textit{\underline{Data availability}-}
The data used to generate and analyze the results may be obtained from the corresponding author upon reasonable request.

\textit{\underline{\smash{Acknowledgments}}-} 
F.A.A. acknowledges the `International Centre for Theory of Quantum Technologies’ project~(contract no. MAB/2018/5) that is carried out within the International Research Agendas Programme of the Foundation for Polish Science financed by the European Union from the funds of the Smart Growth Operational Programme. P.R.D., and M.P. acknowledge the support from the Narodowe Centrum Nauki~(NCN) Poland, \mbox{ChistEra-2023/05/Y/ST2/00005} under the project Modern Device Independent Cryptography~(MoDIC). A.S.H. is funded by RAVEN under the Call: \mbox{HORIZON-CL4-2023-DIGITAL-EMERGING-01}, Project ID:101135787.

\appendix
\section{Appendix A: Single-qubit results} \label{app:singlequbit}

In the following, we present definitions of IBM's \mbox{single-qubit} gates used in the experiments ~\cite{ibmLibrarylatest}.

\begin{table}[hbtp!]
  \centering
  \begin{tabular}{|l|cc|}
  \hline
   \bf{Gate} & \bf{Definition} & \\
    \hline
   P & Applies a phase of $e^{i\theta}$ & \\
   \hline 
   RZ &  \begin{tabular}[c]{@{}l@{}}Rotates the qubit state around \\the z-axis by the given angle.\end{tabular} & \\
   \hline
   RX & \begin{tabular}[c]{@{}l@{}}Rotates the qubit state around \\the x-axis by the given angle.\end{tabular} & \\
   \hline
   RY & \begin{tabular}[c]{@{}l@{}}Rotates the qubit state around \\the y-axis by the given angle.\end{tabular} & \\
   \hline
   $\sqrt{x}$ & \begin{tabular}[c]{@{}l@{}}Creates an equal superposition\\state if the qubit is in the state\\$\ket{0}$, but with a different relative\\phase.\end{tabular} & \\
   \hline
   $\sqrt{x}^\dagger$ & Inverse of the $\sqrt{x}$ gate. & \\
   \hline
   Y & The Pauli Y gate. & \\
   \hline
   U & \begin{tabular}[c]{@{}l@{}}Rotates the qubit state around \\the three-axis by the given angles.
   \end{tabular} & \\
   \hline
  \end{tabular}
  \caption{Definition of single-qubit gates.}
  \label{tab:}
\end{table}

\section{Appendix B: Algorithm}
\label{appx:algo}
The algorithm produces different quantum circuits to design experiments that use \mbox{cross-talk} produced by QG to launch a row hammer attack in order to flip a qubit state as much as possible.

\begin{algorithm}[!h!]
	\caption{Factory}
	\label{al:cross-talk}
	\SetKwInput{KwInput}{Input}
	\KwInput{center=72 extraSets= 0}
	
	\SetKwFunction{FGetNodes}{getNodes}
	\SetKwProg{Fn}{Function}{:}{}
	\Fn{\FGetNodes{node}}{
	\KwRet adjacent nodes}
	
	\SetKwFunction{FGetNodesCombination}{getNodesCombination}
	\SetKwProg{Fn}{Function}{:}{}
	\Fn{\FGetNodesCombination{lst1, lst2}}{
	lstNode=[]\;
	\For{a in lst1}{
		save in lstNode the bidirectional connection among the qubits on the same row\;
		\For{b in lst2}{
			save in lstNode the bidirectional connection among the qubits of the two rows
		}
	}
	\KwRet lstNode or list with the bidirectional connection between the rows' centre and adjacent qubits}
	
	\SetKwFunction{makeCircuitAndMeasure}{makeCircuitAndMeasure}
	\SetKwProg{Fn}{Function}{:}{}
	\Fn{\makeCircuitAndMeasure{centerMeasure,cycles,\\methodName}}{
	channels = getNodes(centerMeasure)\;
	\tcp{Depending on methodName, build the circuit on the channels}
	\ElseIf{methodName == cx}{
		\For{\_ in range(0,cycles)}{
			\For{x in getNodesCombination([channels])}{
				fn(qreg\_q[x[0]], qreg\_q[x[1]])
			}
		qc.barrier(*qreg\_q[0:qNumber])
		}
	}
	qc.measure(qreg\_q[centerMeasure], creg\_c[0])\;
	\KwRet circuit}
	
	circuit1 = makeCircuitAndMeasure(center,1+extraSets,'sx')\;
	expLst.append(transpile(circuit1, backend, optimization\_level=0))\;
	
	\SetKwInput{KwOutput}{Output}
	\KwOutput{CSV file}
\end{algorithm}

\section{Appendix C: Flipping rates of single-qubits}
\label{appx:fliprate}
The following tables show the flipping rates corresponding to each \mbox{single-qubit} gate. The Output $0$ and Output $1$ indicate changes in the measurement outcomes based on whether a bit flip did not occur or occur respectively. Hence, this implies that the experimental results are stable and reliable for using \mbox{cross-talk} to implement the row hammer attack. It can be seen that the rate of bit flip is low for the Kyiv QC as compared to the Osaka QC.

\begin{table}[!h!]
  \centering
    \resizebox{\rt\linewidth}{!}{
\begin{tabular}{rcc}\rowcolor[HTML]{C0C0C0} 
Experiment       & Output 0 \% & Output 1  \% \\
Precision(109)   & 94.7125     & 5.2875       \\\rowcolor[HTML]{EFEFEF}
-p(109-1000) 6   & 94.48       & 5.52         \\
-rz(109-1000) 6  & 94.6575     & 5.3425       \\\rowcolor[HTML]{EFEFEF}
-rx(109-1000) 6  & 93.1325     & 6.8675       \\
-ry(109-1000) 6  & 93.325      & 6.675        \\\rowcolor[HTML]{EFEFEF}
-$\sqrt{x}$(109-1000) 6  & 93.905      & 6.095        \\
-$\sqrt{x}^\dagger$(109-1000) 6 & 94.085      & 5.915        \\\rowcolor[HTML]{EFEFEF}
-y(109-1000) 6   & 94.06       & 5.94         \\
-u(109-1000) 6   & 93.4025     & 6.5975 
\end{tabular}}
  \caption{Kyiv single-qubit experimental results.}
\end{table}

\begin{table}[!h!]
  \centering
    \resizebox{\rt\linewidth}{!}{
\begin{tabular}{rcc}\rowcolor[HTML]{C0C0C0} 
\textbf{Experiment} & \textbf{Output 0 \%}   & \textbf{Output 1  \%} \\
Precision(34)       & 83.3525				 & 16.6475               \\\rowcolor[HTML]{EFEFEF}
p(34-1000) 6        & 82.965                 & 17.035                \\
rz(34-1000) 6       & 17.1025                & 82.8975               \\\rowcolor[HTML]{EFEFEF}
rx(34-1000) 6       & 24.1825                & 75.8175               \\
ry(34-1000) 6       & 75.9025                & 24.0975               \\\rowcolor[HTML]{EFEFEF}
$\sqrt{x}$(34-1000) 6       & 21.4125        & 78.5875               \\
$\sqrt{x}^\dagger$(34-1000) 6      & 20.8825 & 79.1175               \\\rowcolor[HTML]{EFEFEF}
y(34-1000) 6        & 79.24                  & 20.76                 \\
u(34-1000) 6        & 75.9675                & 24.0325
\end{tabular}}
  \caption{Osaka single-qubit experimental results.}
\end{table}

\section{Appendix D: Kyiv and Brisbane results}
\label{appx:results}

The results for QCs Kyiv and Brisbane are shown here. These figures represent the flipping rates for different configurations of the \mbox{C-Not} gate. The flipping of the target qubit from $|0\rangle$ to $|1\rangle$ and vice versa is considered to be successful when the observed flipping rate is greater than $^2/_3$ of the total number of experiments performed.

\begin{figure}[h!btp]
  \centering
  \includegraphics[width=\r\linewidth]{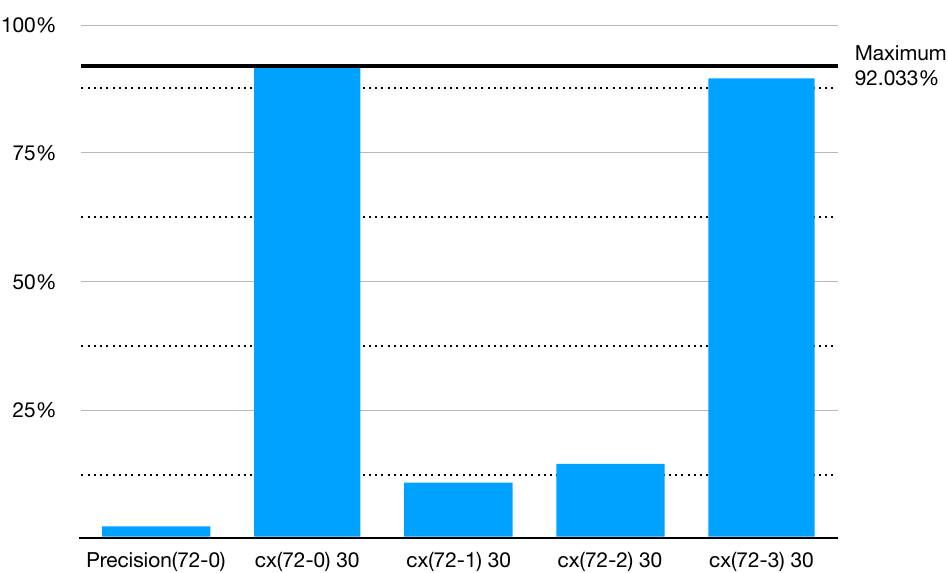}
  \caption{Brisbane A: Flipping $|0\rangle$ to $|1\rangle$. \txt}
\end{figure}

\begin{figure}[h!btp]
  \centering
  \includegraphics[width=\r\linewidth]{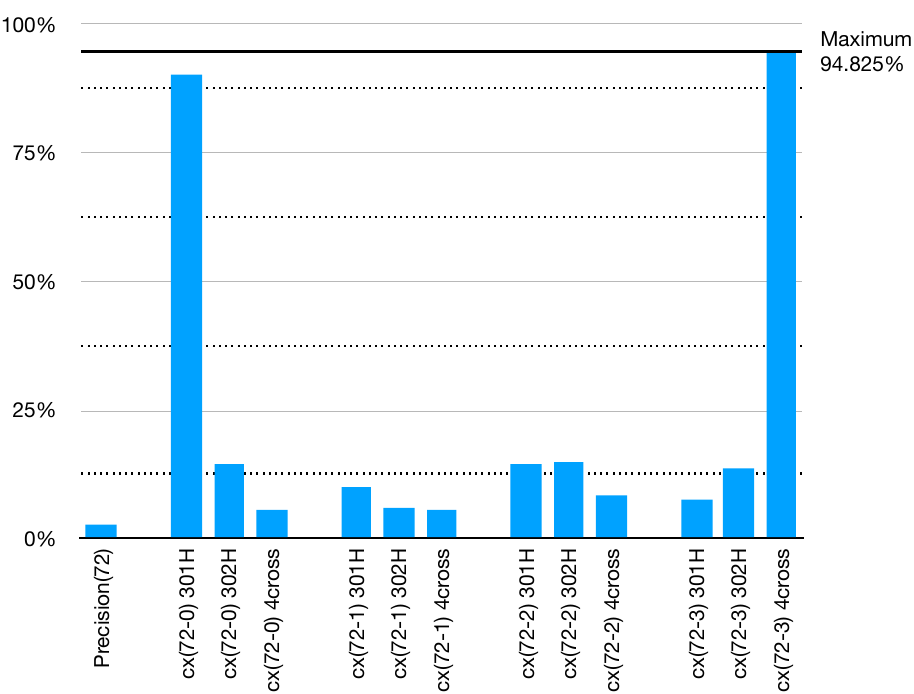}
  \caption{Brisbane B: Flipping $|0\rangle$ to $|1\rangle$. \txt}
\end{figure}

\begin{figure}[h!btp]
  \centering
  \includegraphics[width=\r\linewidth]{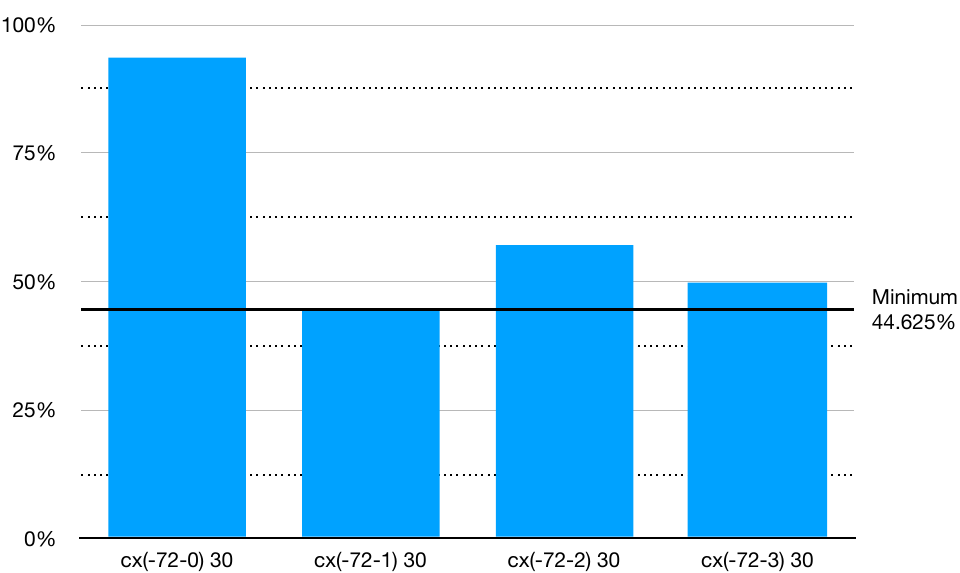}
  \caption{Brisbane: Flipping $|1\rangle$ to $|0\rangle$. \txt}
\end{figure}

\begin{figure}[h!btp]
  \centering
  \includegraphics[width=\r\linewidth]{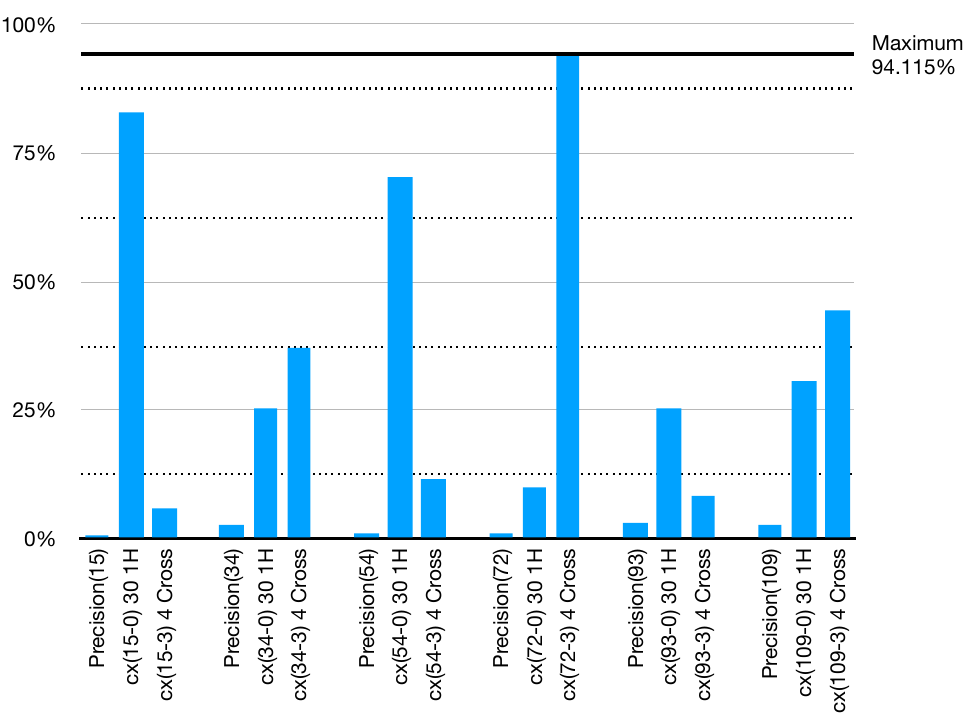}
  \caption{Brisbane: Flipping $|0\rangle$ to $|1\rangle$ on different centres. \txt}
\end{figure}

\begin{figure}[h!btp]
  \centering
  \includegraphics[width=\r\linewidth]{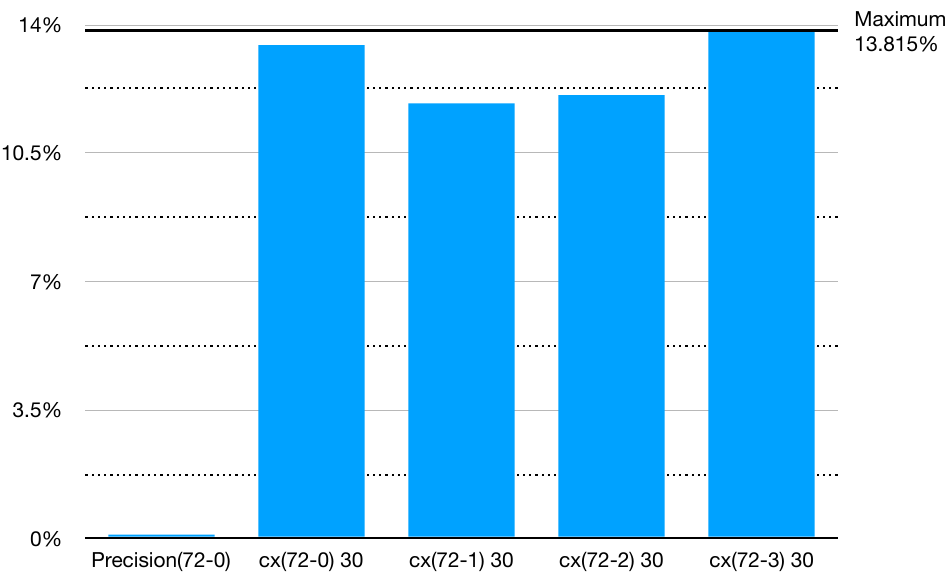}
  \caption{Kyiv A: Flipping $|0\rangle$ to $|1\rangle$. \txt}
\end{figure}

\begin{figure}[h!btp]
  \centering
  \includegraphics[width=\r\linewidth]{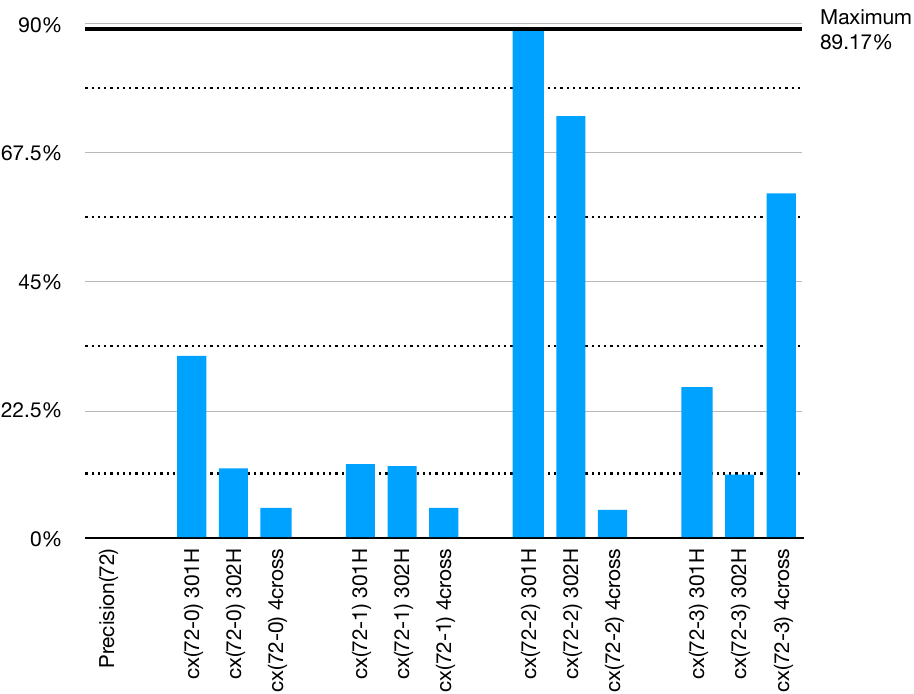}
  \caption{Kyiv B: Flipping $|0\rangle$ to $|1\rangle$. \txt}
\end{figure}

\begin{figure}[h!btp]
  \centering
  \includegraphics[width=\r\linewidth]{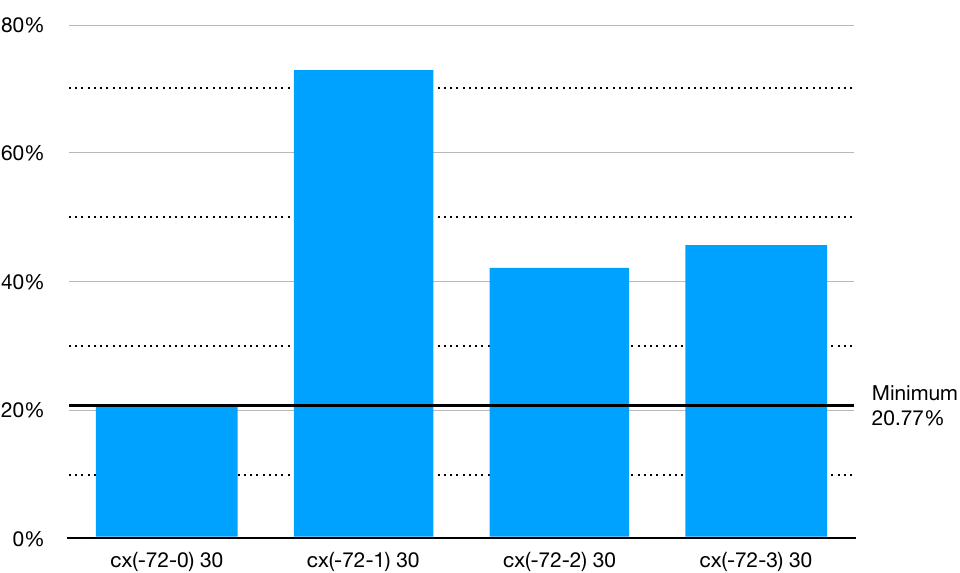}
  \caption{Kyiv: Flipping $|1\rangle$ to $|0\rangle$. \txt}
\end{figure}

\begin{figure}[h!btp]
  \centering
  \includegraphics[width=\r\linewidth]{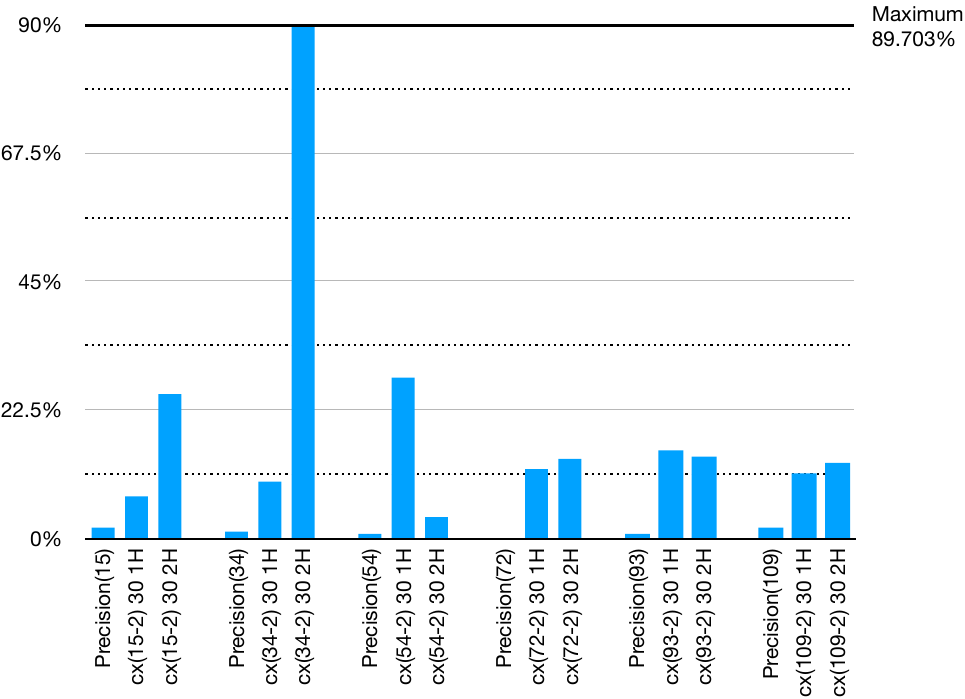}
  \caption{Kyiv: Flipping $|0\rangle$ to $|1\rangle$ states on different centres. \txt}
\end{figure}

\begin{figure}[h!btp]
  \centering
  \includegraphics[width=\r\linewidth]{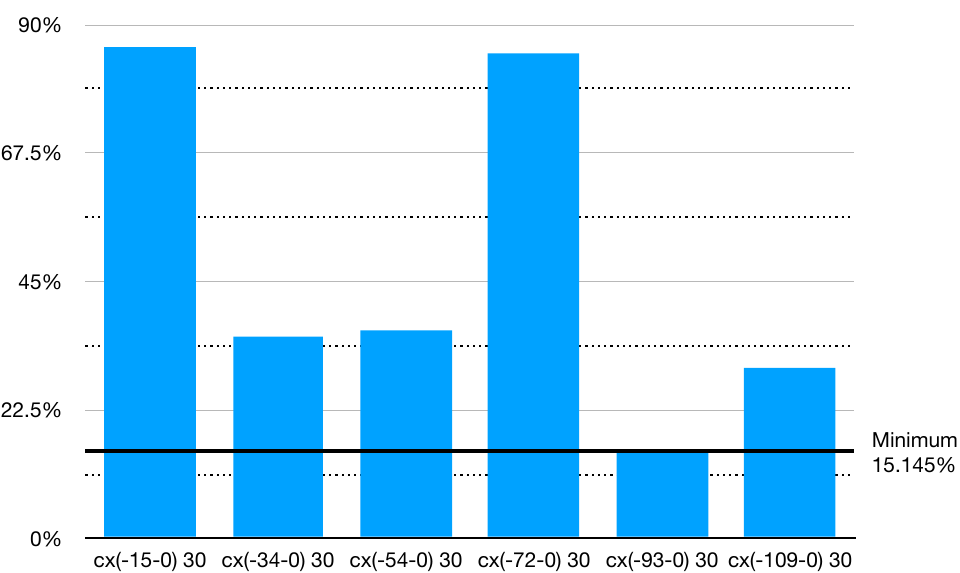}
  \caption{Kyiv: Flipping $\ket{1}$ to $\ket{0}$ states on different centres. \txt}
\end{figure}

\section{Appendix E: Cramér's V}
\label{appx:cramers}

Cramér’s V is an effect size measurement for the \mbox{chi-square} test of independence. It measures how strongly two categorical fields are associated~\cite{ibmCramer}, and is defined as
\begin{equation*}
  \text{Cramér's~V} = \sqrt{\frac{\chi^2}{n\times \text{min}(c-1,\ r-1)}},
  \label{eq:cramersveq}
\end{equation*}
where
\begin{itemize}
  \item $\chi^2$ is the \mbox{Chi-square} statistic,
  \item $n$: is the total sample size,
  \item $r$: is the number of rows in the data,
  \item $c$: is the number of columns in the data.
\end{itemize}
This measure has the following meaning
\begin{table}[h!]
  \centering
  \begin{tabular}{|c|l|}
  \hline
   \bf{Effect size} (\textbf{ES}) & \bf{Interpretation} \\
    \hline
ES = 0                                     & No association among the fields.                                                                                \\
\hline
ES $\le$ 0.2                               & \begin{tabular}[c]{@{}l@{}}The result is weak. Although\\ the result is statistically significant.\end{tabular} \\ \hline
0.2 $<$ ES $\le$ 0.6                       & The fields are moderately associated.                                                                           \\ \hline
ES $>$ 0.6                                 & The fields are strongly associated.                                                                             \\ \hline
ES = 1                                     & \begin{tabular}[c]{@{}l@{}}There is a perfect association\\ among the fields.\end{tabular}  \\  \hline                 
  \end{tabular}
  \caption{Cramér's~V interpretation}
  \label{tab:cramersvint}
\end{table}

\begin{widetext}
	\begin{table}[h!]
\centering
\begin{tabular}{llllllllll}\rowcolor[HTML]{C0C0C0} 
Experiment  & \begin{tabular}[c]{@{}l@{}}Output 0\\frequency\end{tabular} & \begin{tabular}[c]{@{}l@{}}Output 1\\frequency\end{tabular} & Total  & Expected$_0$ & Expected$_1$ & $\left[^{(O-E)^2}/_E\right]_0$ & $\left[^{(O-E)^2}/_E\right]_1$ & Label & Result \\
Precision(72)   & 38925              & 1075               & 40000  & 27444.41 & 12555.58 & 4802.57 & 10497.62 & $\chi^2$           & 334515.18 \\\rowcolor[HTML]{EFEFEF}
cx(-72-0) 30    & 2644               & 37356              & 40000  & 27444.41 & 12555.58 & 22411.13 & 48986.98 & min (c-1,r-1) & 1                \\
cx(72-0) 301H   & 3912               & 36088              & 40000  & 27444.41 & 12555.58 & 20178.03 & 44105.81 & n            & 680000           \\\rowcolor[HTML]{EFEFEF}
cx(72-0) 302H   & 34282              & 5718               & 40000  & 27444.41 & 12555.58 & 1703.53 & 3723.64 & Cramér’s V   & 0.70 \\
cx(72-0) 4cross & 37728              & 2272               & 40000  & 27444.41 & 12555.58 & 3853.32 & 8422.71 &              &                  \\\rowcolor[HTML]{EFEFEF}
cx(-72-1) 30    & 22150              & 17850              & 40000  & 27444.41 & 12555.58 & 1021.36  & 2232.53 &              &                  \\
cx(72-1) 301H   & 36013              & 3987               & 40000  & 27444.41 & 12555.58 & 2675.25 & 5847.65  &              &                  \\\rowcolor[HTML]{EFEFEF}
cx(72-1) 302H   & 37547              & 2453               & 40000  & 27444.41 & 12555.58 & 3718.87 & 8128.83 &              &                  \\
cx(72-1) 4cross & 37795              & 2205               & 40000  & 27444.41 & 12555.58 & 3903.69 & 8532.82  &              &                  \\\rowcolor[HTML]{EFEFEF}
cx(-72-2) 30    & 17153              & 22847              & 40000  & 27444.41 & 12555.58 & 3859.18   & 8435.53 &              &                  \\
cx(72-2) 301H   & 34160              & 5840               & 40000  & 27444.41 & 12555.58 & 1643.28 & 3591.95 &              &                  \\\rowcolor[HTML]{EFEFEF}
cx(72-2) 302H   & 34086              & 5914               & 40000  & 27444.41 & 12555.58 & 1607.27 & 3513.23 &              &                  \\
cx(72-2) 4cross & 36576              & 3424               & 40000  & 27444.41 & 12555.58 & 3038.35 & 6641.33 &              &                  \\\rowcolor[HTML]{EFEFEF}
cx(-72-3) 30    & 20036              & 19964              & 40000  & 27444.41 & 12555.58 & 1999.84 & 4371.32 &              &                  \\
cx(72-3) 301H   & 36972              & 3028               & 40000  & 27444.41 & 12555.58 & 3307.59 & 7229.84 &              &                  \\\rowcolor[HTML]{EFEFEF}
cx(72-3) 302H   & 34506              & 5494               & 40000  & 27444.41 & 12555.58 & 1816.98 & 3971.62 &              &                  \\
cx(72-3) 4cross & 2070               & 37930              & 40000  & 27444.41 & 12555.58 & 23460.54 & 51280.81 &              &                  \\\rowcolor[HTML]{EFEFEF}
Total           & 466555             & 213445             & 680000 &                  &                  &                  &                  &              &                 
\end{tabular}
	\caption{Cramér’s V Brisbane shows a 0.7 value, which means that these fields, the use of \mbox{C-Not} gates and the \mbox{cross-talk} measured by the qubit flipping are strongly associated.}
\end{table}
\end{widetext}

\bibliographystyle{unsrt}
\bibliography{References}
\end{document}